# Label-free optical quantification of structural alterations in Alzheimer's disease

Moosung Lee[a+], Eeksung Lee[b,d+], JaeHwang Jung[a], Hyeonseung Yu[a], Kyoohyun Kim[a], Shinhwa Lee[c], Yong Jeong[d,*], YongKeun Park[a,e*]

[a] Department of Physics, Korea Advanced Institute of Science and Technology, Daejeon, South Korea, 34141

[b] Graduate School of Medical Science and Engineering, Korea Advanced Institute of Science and Technology, Daejeon 34141, South Korea

[c] Department of Biological Sciences, Korea Advanced Institute of Science and Technology, Daejeon 34141, South Korea

[d] Department of Bio and Brain Engineering, Korea Advanced Institute of Science and Technology, Daejeon 34141, South Korea

[e] TOMOCUBE, Inc., Daejeon 34051, Republic of Korea

+these authors contribute equally to this work.
*corresponding authors: Y.K.P (yk.park@kaist.ac.kr); Y.J (yong@kaist.ac.kr)

**We present a quantitative label-free imaging of mouse whole brain tissue slices with sub-micrometre resolution, employing holographic microscopy and an automated scanning platform. From the measured light field images, scattering coefficients and anisotropies are quantitatively retrieved, which enables access to structural information about brain tissues. As a proof of principle, we demonstrate that these scattering parameters enable us to quantitatively address structural alteration in the brain tissues of mice with Alzheimer's disease.**

Imaging brain tissues is an essential tool in neuroscience because understanding brain structure provides rich information about brain functions and alterations associated with diseases. Conventional imaging techniques include magnetic resonance imaging (MRI) and positron emission tomography (PET), but such techniques are limited by low spatial resolution (around 100 μm[1]). To assess detailed structural information and to determine specific sites for assured diagnosis, histology has complemented MRI and PET. The traditional histological method, however, relies on staining procedures, which require dyeing with intensive labor, provide only local morphology, and distort structural information due to irregular dye binding or sample damage[2].

To overcome these limitations of the staining procedures, label-free imaging techniques have been employed in brain imaging. For example, second harmonic generation was used for non-invasive brain imaging in cellular levels[3]. Raman scattering was also applied to image myelin fibres and amyloid plaques[4] in brains. Optical coherence microscopy (OCM) has also been utilised for label-free histology of a brain tissue[5]. However, they only provide limited qualitative information about tissue structures.

Here, we present a quantitative label-free approach for the investigation of brain tissue structures. Employing quantitative phase imaging (QPI) techniques[6, 7], holographic images (amplitude and phase delay maps) of whole mouse brain tissues are measured with sub-micrometre resolution. This allows us to perform multi-scale imaging of a whole mouse brain tissue slice, which covers the sub-micrometre scale (subcellular organelles) to the millimetre scale (histological anatomy). We also present a modified version of scattering-phase theorem in order to precisely retrieve scattering coefficients ($\mu_s$) and anisotropies ($g$) maps of tissue slices from the measured holograms, which allows us to investigate the structural organisations of tissues quantitatively. As a proof of principle, we demonstrate that the scattering parameters of brain tissues of mice with Alzheimer's disease (AD) are significantly modified, suggesting that the present approach provides a unique means to investigate pathophysiology of neurological disorders.

The schematic of the QPI setup is shown in Fig. 1a. To perform multi-scale QPI, diffraction phase microscopy (DPM) equipped with a motorised sample stage is utilised (see *Methods*). To measure QPI images of a whole mouse brain slice, segmented QPI images are measured with the sub-micrometre resolution and digitally stitched to generate the image of the whole brain tissue slice. The reconstructed image of a representative brain tissue is presented in Figs. 1f–h. The total field of view is 8.9 × 6.6 mm (horizontal × vertical) and the lateral resolution is 0.8 μm, which is limited only by the numerical aperture (NA) of DPM in this study. The corresponding bright-field images are also presented (Figs. 1i–k). With high contrast and transverse resolution, the phase images show the general anatomical features in a brain and spatial variations at subcellular structure level, these being undistinguishable with bright-field imaging. The images of the adjacent slice stained with hematoxylin and eosin (H&E) are also presented in Figs. 1l–n. A side-by-side comparison reveals the matching of general morphological features between the phase image and the conventional histology.

From the measured field images, the maps of $\mu_s$ and $g$ are precisely retrieved. According to the light scattering theory, $\mu_s$ and $g$, which are thickness-independent parameters, quantify the attenuation of unscattered intensity per unit distance and the degree of forward scattering, respectively[8]. For a thin biological tissue, $\mu_s$ and $g$ are related to the fluctuation of the scattered field, including both amplitude and phase delay. Recently, the scattering-phase theorem was introduced to retrieve $\mu_s$ and $g$ from quantitative phase images[9]. To consider the alteration in amplitude in addition to the phase of the optical field that transmitted the tissue slices, we propose a modified version of the scattering-phase theorem that measures accurate values of $\mu_s$ and $g$, as shown in Figs. 2a–b (*see Methods* and *Supplementary Information*). The magnified view around hippocampi and dentate gyri is shown in Figs. 2c–d with the H&E stained micrograph of the adjacent tissue slice for comparison purpose (Fig. 2e).

The maps of scattering parameters ($\mu_s$ and $g$) provide structural distinctions, thus serving as label-free biomarkers. The boundaries between the grey, the white matter, and the hippocampus formation are clearly separated. Other brain regions such as the thalamus and hypothalamus also exhibit distinct boundaries (the yellow dotted lines in Figs. 2c–e). Importantly, dentate gyri are clearly distinguished from the cornu ammonis (CA) due to their high values of $\mu_s$ and $g$ (the cyan lines in Figs. 2c–e).

The distinct distribution of the values $\mu_s$ and $g$ in the brain regions may relate to the different arrangements of cell types and subcellular compositions in the sub-regions of brains. The area with the highest values of $\mu_s$ corresponds to white matter. This is consistent with the fact that there are more lipid contents in white matter than in grey matter and hippocampi[10]. The lipid contents cause high local contrast of refractive index (RI), resulting in high values of $\mu_s$. The dentate gyri exhibit higher scattering coefficients than the other regions in the hippocampi, primarily due to the presence of dense layers of neurons such as granule cells[11].

Over the entire brain tissues, the retrieved map of $g$ exhibits a value range between 0.9 to 1, which is comparable to other types of tissues[8]. Nevertheless, the distribution of $g$ differs in various sub-regions; the larger size of scattering particles results in the increase of $g$. The map of $g$ indicates higher values in the white matter, indicating the tendency of forward-directed light scattering. Together with the retrieved values of $\mu_s$, this result indicates that the white matter consists of tissue components of inhomogeneous and large scattering particles, mostly bulky myelin sheaths. The areas of grey matter and hippocampi exhibit lower values of $\mu_s$ and $g$, implying that these areas are composed of tissue components of uniformly packed small scattering particles, mostly neuronal cells.

To demonstrate the applicability of the present approach, we have systematically compared the brain tissue slices of AD model mice with their wild-type littermates. We utilised the maps of $\mu_s$ and $g$ for addressing structural alterations in brain tissues associated with AD. Brain tissue slices of the same anatomical regions were measured in five mice from each group. Figures 3a–b show the representative maps for AD and wild-type mice, which show that both the values of $\mu_s$ and $g$ generally increase throughout grey matter and hippocampi in the AD model.

For a quantitative analysis, we statistically analysed the distributions of $\mu_s$ and $g$ in three distinct regions in the brain: grey, white matter, and hippocampi, as shown in Fig. 3c. (Numerical values are shown in **Supplementary Table 2** and **3**). A comparison of sample-mean distribution indicates significant increases of $\mu_s$ and $g$ in the grey and hippocampal regions of the AD models, suggesting that AD pathology alters tissues in these areas to exhibit higher RI inhomogeneity and increased size of scattering particles. These results are consistent with the fact that AD is characterised by the accumulation of extracellular amyloid and intracellular neurofibrillary tangles with larger size than granule cells[12], as well as neuronal cell losses throughout grey matter and hippocampi[13]. The structural changes by AD pathology are mostly distributed over the grey matter and hippocampi, while white matter is relatively spared.

To further investigate the structural alterations in the brain tissues associated with AD, correlative analysis about scattering parameters was performed (Fig. 3d). Figure 3d separately shows the scatterplots of $\mu_s$ and $g$ for the grey matter, white matter and hippocampi in each group. While the scatterplots exhibit a similar correlation of $\mu_s$ and $g$ in white matters for the AD and wild-type mice, they showed a significant difference in the grey matter and hippocampi between the AD and wild-type mice. These trends are clearly visualised when the 50% density boundaries of the scatterplots of $\mu_s$ and $g$ were simultaneously displayed for the six different populations (Fig. 3e).

We present the wide-field label-free QPI of whole mouse brain tissue slices. We demonstrate the multi-scale phase images of the brain tissues with an image-stitching scheme and the accurate retrieval of the scattering parameters with the modified scattering theorem. With AD model mice, we suggest that this approach can be used to systematically quantify the structural changes of AD pathology by measuring quantitative phase images of brain tissue slices. The retrieval of scattering parameters eliminates staining procedures and also provides the quantitative information of the morphology that H&E histology cannot. Although this work has focused on brain tissue slices, the approach is sufficiently broad and general, and it will directly offer novel approaches for general histopathology.

**METHODS**

Methods and any associated references are available in the online version of the paper.

Note: Any Supplementary Information and Source Data files are available in the online version of the paper.

**ACKNOWLEDGEMENTS**

This work was supported by KAIST, and the National Research Foundation of Korea (2015R1A3A2066550, 2014K1A3A1A09063027, 2012-M3C1A1-048860, 2014M3C1A3052537) and Innopolis foundation (A2015DD126).

**AUTHOR CONTRIBUTIONS**

Y.-K.P. and Y.J. initiated the work and supervised the project. M.L. developed the method. M.L., E.L., J.-H.J., H.Y., K.K., and S.L. performed experiments. All the authors wrote the manuscript.

**COMPETING FINANCIAL INTERESTS**

The authors declare no competing financial interests

## ONLINE METHODS

### Optical setup of diffraction phase microscopy

In order to obtain the optical field images of brain tissue slices, DPM has been employed. DPM is a common-path interferometric microscopy, capable of quantitative phase imaging of biological samples with high resolution and phase stability[14]. As an illumination source, a diode-pumped solid-state laser ($\lambda$ = 520 nm, LP520-SF15, Thorlabs Inc., USA) is employed. An inverted microscope (IX71, Olympus American Inc., USA) is equipped with a motorised *xy*-scanning stage (MLS203, Thorlab Inc., USA) and a 10× objective lens (0.4 NA, UPlanSApo 10×, Olympus American Inc., USA). Via the objective and a tube lens, the optical field of a sample is imaged onto the image plane. To acquire a hologram, a grating (92 grooves/mm, 46-071, Edmund Optics Inc., USA) is placed at the sample plane, and a customised pinhole filter is located on the Fourier plane. The diffraction grating splits the sample field into many diffraction orders. Among them, only two orders (the 0th and the 1st diffraction order) of the diffracted beams are further projected onto a CCD plane via a 4-*f* telescopic imaging system. The 1st diffraction beam is spatially filtered using the pinhole (25 μm diameter) located at the Fourier plane, which serves as a reference beam.

Both the sample and the reference beams interferes at the CCD plane, generating a spatially modulated hologram (Fig. 1b). A total magnification of the imaging system is 71×. Individual holograms were recorded by a sCMOS camera (C11440-22C, Hamamatsu Inc., Japan). Using the field retrieval algorithm[15], the amplitude and the phase delay maps of a sample are retrieved (Figs. 1c–d). The amplitude image (Fig. 1c) does not exhibit enough contrast to distinguish geometric anatomy and cellular components due to the optical transparency of tissues. The phase image of the tissue (Fig. 1d), however, clearly shows the meso-scale and the sub-micrometre scale structures of the tissue with significant phase imaging contrast. The resolution of the imaging system is 0.8 μm, which is limited by the numerical aperture of the objective lens (NA = 0.4).

### Stitching segmented quantitative phase images

The full-field phase image of a mouse brain tissue is reconstructed from the measured segmented phase images by stitching the segmented images. In order to achieve wide-field QPI, the motorised stage laterally translates a tissue slice in a fully automated manner. The field of view of a single mosaic image is 187 × 187 μm, which overlaps 10% with adjacent images. A custom-made MATLAB code was used to reconstruct a full-field phase image from more than 1000 mosaics, i.e. 1908 mosaics for Fig. 1f (53 × 36).

The reconstruction includes several imaging processes in order to correctly stitch individual segmented phase images. First, the static phase distortion due to the aberration of the imaging system is removed by subtracting mosaic images by the background phase image. Remaining phase error was reduced by matching the overlapped adjacent areas between mosaic images. A constant phase value is added to the reconstructed image so that the background area has the phase value of zero. Both amplitude and phase images are corrected similarly.

### Sample preparation

For the preparation of brain tissue slices, we used 22-month-old male AβPP$^{SWE}$/PS1ΔE9 transgenic mice (Tg) as AD models and their wild-type (Wt) littermates as control models. The mice were deeply anaesthetised by intraperitoneal injection of ketamine/xylazine and perfused transcardially with phosphate buffered saline (PBS), then with a fixative containing 4% paraformaldehyde in a phosphate buffered saline solution. The brains were embedded in paraffin block and sliced into 3–3.5 μm-thick coronal sections using microtome (RM2245, Leica Microsystems, Germany). The sample preparation procedures and the methods were reviewed and approved by the Institutional Review Board (KA-2015-03).

### Accurate retrieval of $\mu_s$ and $g$

We utilised the modified scattering theorem to retrieve the maps of $\mu_s$ and $g$ over the entire brain tissue with the sub-region window as 18 μm × 18 μm. This dimension was both large enough to represent local structural variations and small enough to correspond to the cellular size. $\mu_s$ is accurately retrieved by Beer-Lambert law,

$$\mu_s = \frac{\ln(I_0 / I)}{L}, \quad (1)$$

where $L$ is the thickness of a tissue, $I$ the unscattered light intensity and $I_0$ the total illumination intensity, which is the sum of scattered intensity and unscattered intensity. The scattered intensity map is retrieved from the measured field by the following relation:

$$I(\mathbf{q}) = \frac{1}{2\pi}\left|\int U(\mathbf{r})e^{-i\mathbf{q}\cdot\mathbf{r}}d^2\mathbf{r}\right|^2, (2)$$

, where $\mathbf{q}$ is the spatial frequency vector and $U(\mathbf{r}) = A(\mathbf{r})\exp[i\phi(\mathbf{r})]$ is the measured complex optical field by DPM, $A(\mathbf{r})$ and $\phi(\mathbf{r})$ are the amplitude and the phase of the field. $g$ was computed by the following relation:

$$g = 1 - \frac{1}{2k_0^2}\left\langle |\nabla[U(\mathbf{r})]^N|^2 \right\rangle_\mathbf{r} = 1 - \frac{1}{2k_0^2}\left\langle |\nabla A_{l_s}(\mathbf{r})|^2 + \left(\frac{l_s}{L}\right)^2 |A_{l_s}(\mathbf{r})\nabla\phi(\mathbf{r})|^2 \right\rangle_\mathbf{r}, (3)$$

where $l_s = 1/\mu_s$ is mean free path, $N = l_s / L$ is the ratio of mean free path to the thickness of a sample, $A_{ls}(\mathbf{r}) = A(\mathbf{r})^N$ is expected amplitude when light passes through the sample whose thickness is $l_s$, and $\langle\ \rangle_\mathbf{r}$ denotes the spatial average value over a sub-region window. More explanation is shown in *Supplementary Information*.

**Statistical Analysis.** *P* values were calculated by Mann-Whitney-Wilcoxon rank tests comparing the sample means of scattering coefficients and anisotropy values between normal and AD models. All of the numbers following the ± sign in the text are standard deviations.

Figures

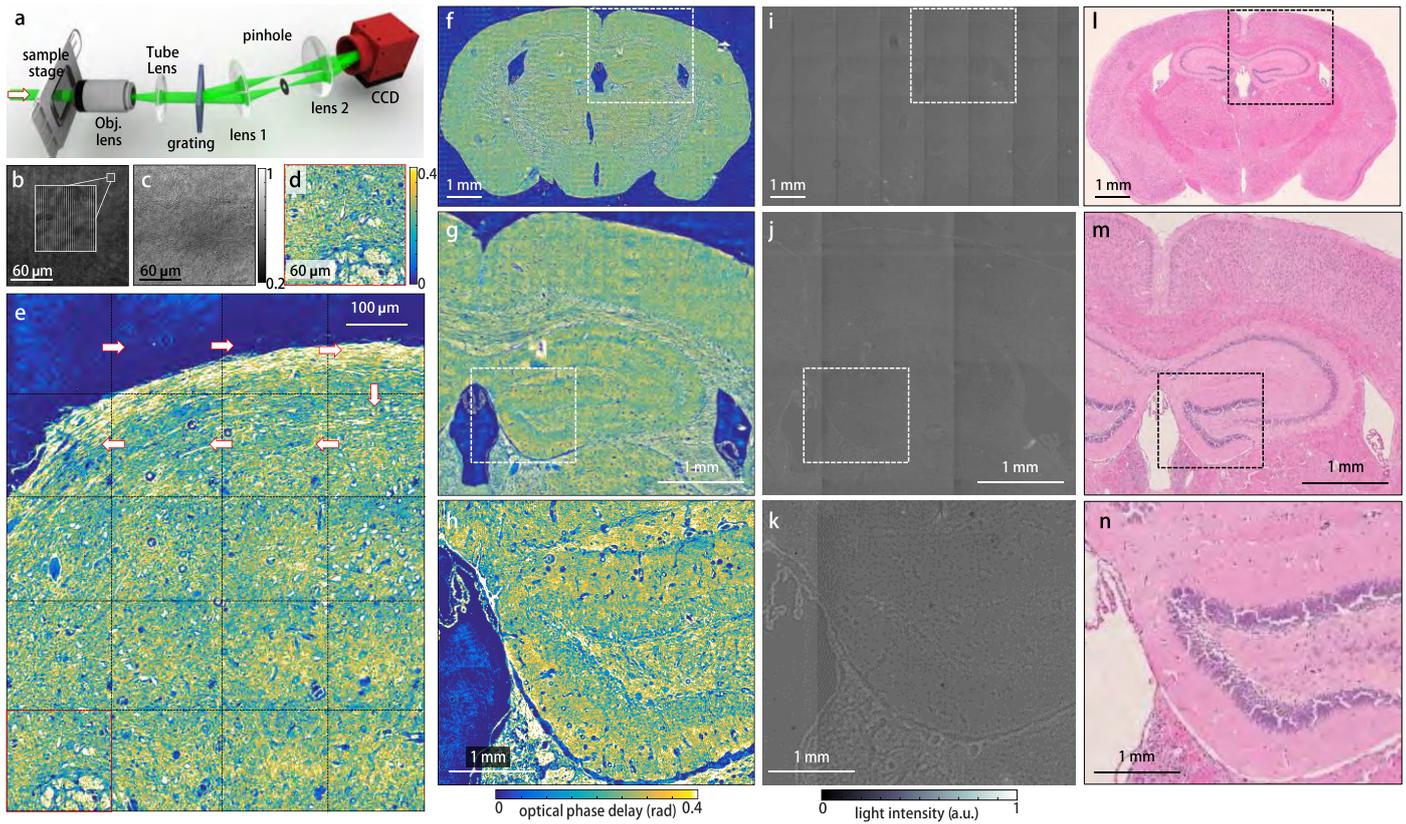

**Figure 1.** Quantitative phase imaging of a mouse brain tissue slice. **(a)** Diffraction phase microscopy equipped with a translational stage for measuring optical phase delays. **(b)** A representative hologram. **(c)** The retrieved amplitude and **(d)** phase image retrieved from the hologram in (b). **(e)** Wide-field phase images of the mouse brain slice, stitched from individual holograms (dotted boxes). Arrows indicate a recording order. **(f-h)** Phase delay image measured with DPM. **(i-k)** Bright-field image of an unstained brain tissue **(l-n)** H&E stained brain tissue slice micrograph of the same brain tissue. For each imaging modality, the magnified image of a selected region in the dashed square is represented below.

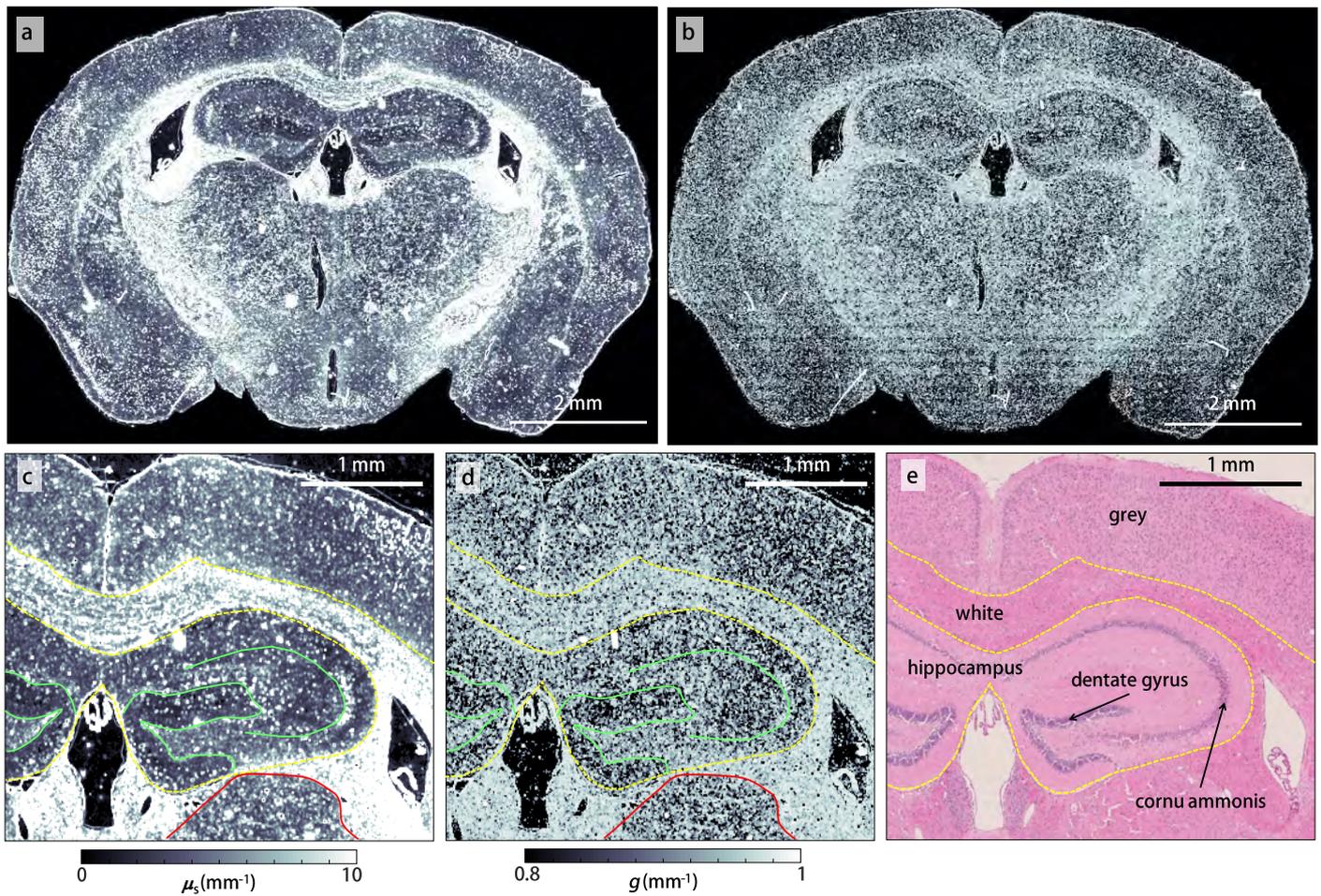

**Figure 2. Scattering parameter maps of the brain tissue slice obtained from the quantitative phase image. (a)** Full-field maps of scattering coefficient and **(b)** anisotropy obtained from the quantitative phase images with the modified scattering theorem. **(c)** Magnified images of scattering coefficient map and **(d)** anisotropy map. **(e)** A matched H&E stained image with **(c-d)**. Boundaries between grey matter, white matter, and hippocampus are indicated with the dotted yellow lines. Boundaries of dentate gyrus and cornu ammonis are presented with the cyan lines

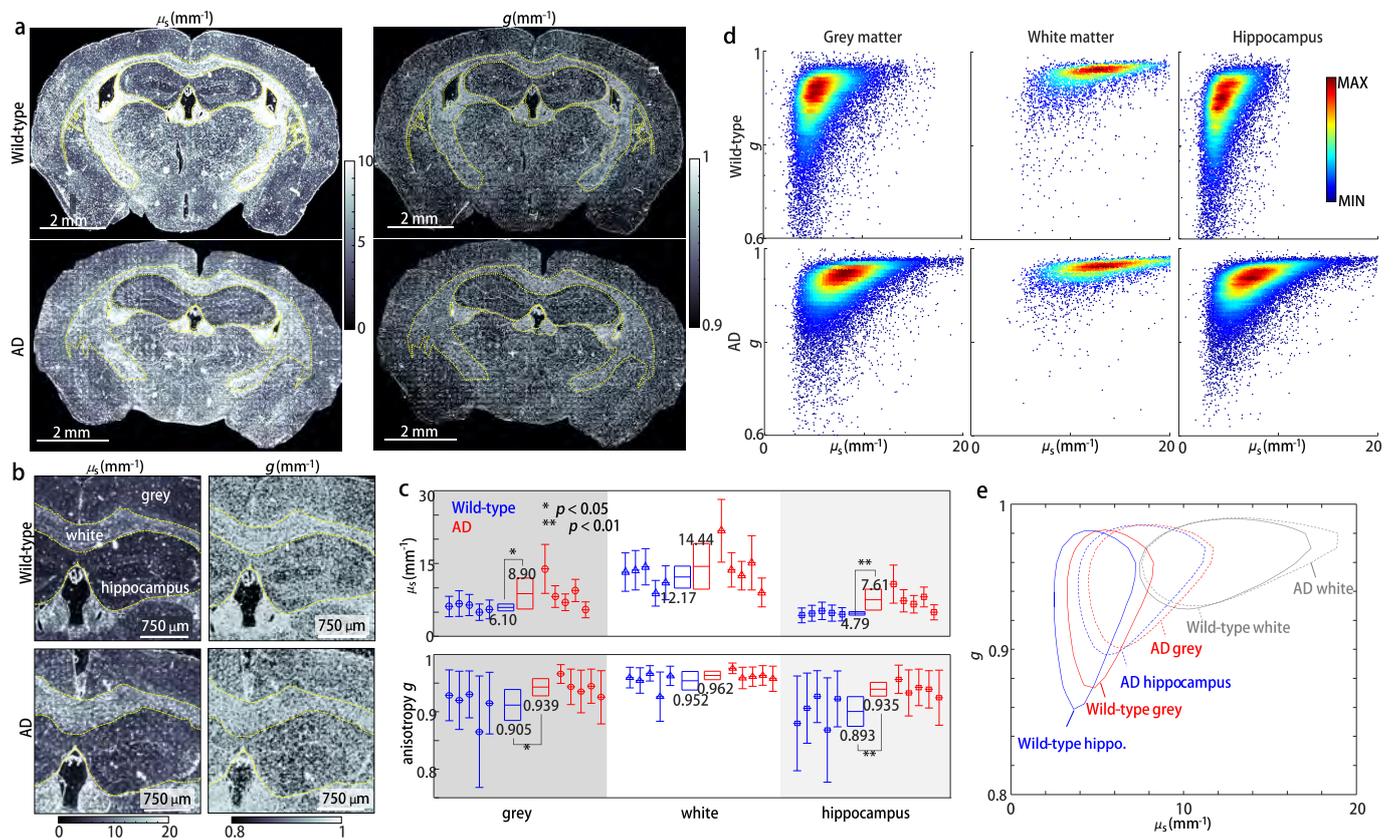

**Figure 3. Quantitative analysis of scattering parameters in brain tissues and their alterations in Alzheimer's disease model mice. (a)** Full-field maps of scattering coefficient and anisotropy of representative AD and wild-type mouse. **(b)** Magnified images of scattering parameter maps in (a), showing clear distinction between grey matter, white matter, and hippocampi. **(c)** Distributions of scattering coefficients and anisotropy values in grey matter, white matter, and hippocampus regions for five mice from each model. Linear bar: distribution of scattering parameters in each different tissue. Rectangular bar: Sample-mean distribution of the scattering parameters in each sub-region. Range of the bar is mean ± standard deviation. *: $p < 0.05$, **: $p < 0.01$. **(d)** Probability density maps of the scattering parameters in grey matter, white matter, and hippocampus for healthy and Alzheimer's disease mice. **(e)** 50% density contour plots of the six scatterplots in (d).